\documentclass[12pt, letterpaper, twoside]{article}
\usepackage[utf8]{inputenc}
\usepackage{graphicx}
\usepackage[colorlinks, citecolor = blue, linkcolor = blue, urlcolor=blue]{hyperref}
\usepackage{mathtools}
\usepackage{amsmath}
\usepackage{amssymb}
\usepackage{alphabeta}
\usepackage{systeme}
\usepackage[table]{xcolor}
\usepackage{fancyhdr}
\usepackage{subfigure}
\usepackage{setspace}
\usepackage{cleveref}
\usepackage{amsmath}
\usepackage{tikz}
\usepackage{newclude}
\usepackage{ragged2e}
\usepackage[ruled,vlined]{algorithm2e}
\usepackage{algorithmic}
\newcommand\numberthis{\addtocounter{equation}{1}\tag{\theequation}}
\usepackage{comment}
\usepackage{ulem}

\usepackage[margin=1in]{geometry}

\usepackage[
    backend=biber,
    style=chem-angew,
  ]{biblatex}
\usepackage{authblk}
\date{}
\title{\textbf{An MCMC Approach to Bayesian Image Analysis in Fourier Space}}

\author[1]{Konstantinos Bakas}
\author[2]{John Kornak}
\author[1]{Hernando Ombao}
\affil[1]{\small King Abdullah University of Science and Technology, SA}
\affil[2]{\small University of California, San Francisco, USA} 
\affil[ ]{konstantinos.bakas@kaust.edu.sa, john.kornak@ucsf.edu, hernando.ombao@kaust.edu.sa}

\begin{document}
\maketitle

\begin{abstract}
Bayesian methods are commonly applied to solve image analysis problems such as noise-reduction, feature enhancement and object detection. A primary limitation of these approaches is the computational complexity due to the interdependence of neighboring pixels which limits the ability to perform full posterior sampling through Markov chain Monte Carlo (MCMC). To alleviate this problem, we develop a new posterior sampling method that is based on modeling the prior and likelihood in the space of the Fourier transform of the image. One advantage of Fourier-based methods is that many spatially correlated processes in image space can be represented via independent
processes over Fourier space. A recent approach known as Bayesian Image Analysis in Fourier Space (or BIFS), has introduced parameter functions to describe prior expectations about image properties in Fourier space. To date BIFS has relied on Maximum a Posteriori (MAP) estimation for generating posterior estimates; providing just a single point estimate. The work presented here develops a posterior sampling approach for BIFS  that can explore the full posterior distribution while continuing to take advantage of the independence modeling over Fourier space. As a result
computational efficiency is improved over that for conventional Bayesian image analysis and mixing concerns that commonly have to be dealt with in high dimensional Markov chain Monte Carlo sampling problems are avoided. Implementation results and details are provided using simulated data.
\end{abstract}

\section{Introduction}\label{Introduction}

An image is worth a thousand words. However, the presence of noise can negatively affect the information that an image carries. Bayesian image analysis methods try to tackle this issue by improving image quality and enhancing important characteristics of the image. Applying Bayes Theorem, the reconstruction (posterior) comes from combining a priori expectations of the spatial characteristics of the image (prior) with the noise degradation process (likelihood). Suppose that $x$ is the true image (noise-free) and $y$ is the sub-optimal one (exposed to noise). Our goal is to estimate $x$. The posterior distribution of $x$, $\pi(x|y)$, is proportional to the product of the likelihood $\pi(y|x)$ and prior $\pi(x)$, that is 
\begin{equation}
    \pi(x|y) \propto \pi(y|x) \pi(x).
\end{equation}
\par The majority of conventional Bayesian imaging models apply Bayes Theorem in \emph{image space}, i.e., in terms of pixel elements in 2D Euclidean space. Typically the most critical component for the reconstruction is the choice of the prior distribution. One of the most widely used models for the prior is a Markov Random Field (MRF) \cite{besag1989digital}, \cite{geman1984stochastic}. A MRF consists of a set of random variables which obey the Markov property and the random variables ``communicate'' with each other via an undirected graph. We  denote as $x_s$ the random variable associated with the node (or pixel) $s \in S$ and as $\partial x_s$ its neighborhood region. Then the Markov property implies that the distribution of $x_s$ is independent of the process outside of its neighborhood $\partial x_s$, that is
\begin{equation}
    \pi(x_s | x_{-s}) = \pi(x_s | \partial x_s).
\end{equation}
\par The inter-dependencies between the pixels of an image form a distribution in their neighborhood. It is apparent that neighboring pixels are more closely related, than pixels that are far apart. Certainly, the range of the neighborhood structure may vary depending on the prior characteristics to be defined. Therefore, defining an appropriate neighborhood structure and parameter values is quite difficult under the setting of MRF models. The definition of an MRF over a set of random variables $S$ is given via the conditional probability density function of individual pixel intensity values given its neighborhood region. Then the posterior distribution of $x_s$ given the rest of the pixels in the image $x_{-s}$ and the observed data $y$ can be written as,
\begin{equation}
    \pi(x_s|y, x_{-s}) \propto \pi(y_s|x_s) \pi(x_s|\partial x_s).
    \label{eq1}
\end{equation}
\par Of course, the choice of the neighborhood structure as well as the weights they have on the pixel of interest need to be considered with caution. There is a trade-off between the size of the neighborhood region and the computational expedience of the model. Moreover, the bias and variance problem has to be faced as well. If the neighboring region is set to be broad then this could introduce high bias whereas if it is set to be small that could induce the variance to decay to zero significantly slowly. The ideal situation is for the neighborhood of $x_s$ to be small enough to capture the most useful information about $x_s$ without unnecessarily increasing
the bias and computational cost. In this way, the precision matrix of the prior will be sparse. Therefore, only a small amount of computational power will be required for the reconstruction of the image. The definition of these priors has an intuitive interpretation. However, these models come with some drawbacks. The main disadvantages of those methods are (a) the slow computational speed and (b) the code is not robust to the distributional structure of the neighborhood (change at the neighborhood structure will require a new MCMC code). However, we can overcome such problems in Fourier space.
\par More recently, a new method has been proposed in \cite{kornak2020bayesian}, named Bayesian Image analysis in Fourier Space (BIFS), which transforms the Bayesian image analysis model from image space into Fourier space utilizing the two-dimensional Discrete Fourier Transform (DFT). We define $f$ to be a two dimensional spatial function (i.e., image) and $F$ to be its representation under the Fourier basis system. Note that there is 1-1 mapping between the spatial domain signal and the frequency domain signal. In particularly, the two dimensional transformation from image space to Fourier space is given by:
\begin{equation}
    \mathcal{F}(u,v) = \frac{1}{MN} \sum_{m=0}^{M-1} \sum_{n=0}^{N-1} f(m,n) \exp\bigl (-i(um + vn) \bigr)
    \label{dft}
\end{equation}
\noindent where, $M\times N$ is the dimension of the spatial signal. To calculate the DFT the Fast Fourier Transform (FFT) algorithm is used for computational expedience. FFT reduces the complexity of calculating Equation \ref{dft} from $\mathcal{O}((MN)^2)$ to $\mathcal{O}((MN)\log(MN))$ \cite{cooley1969fast}.
\par
\begin{figure}
    \centering
    \includegraphics[width=15cm, height=5cm]{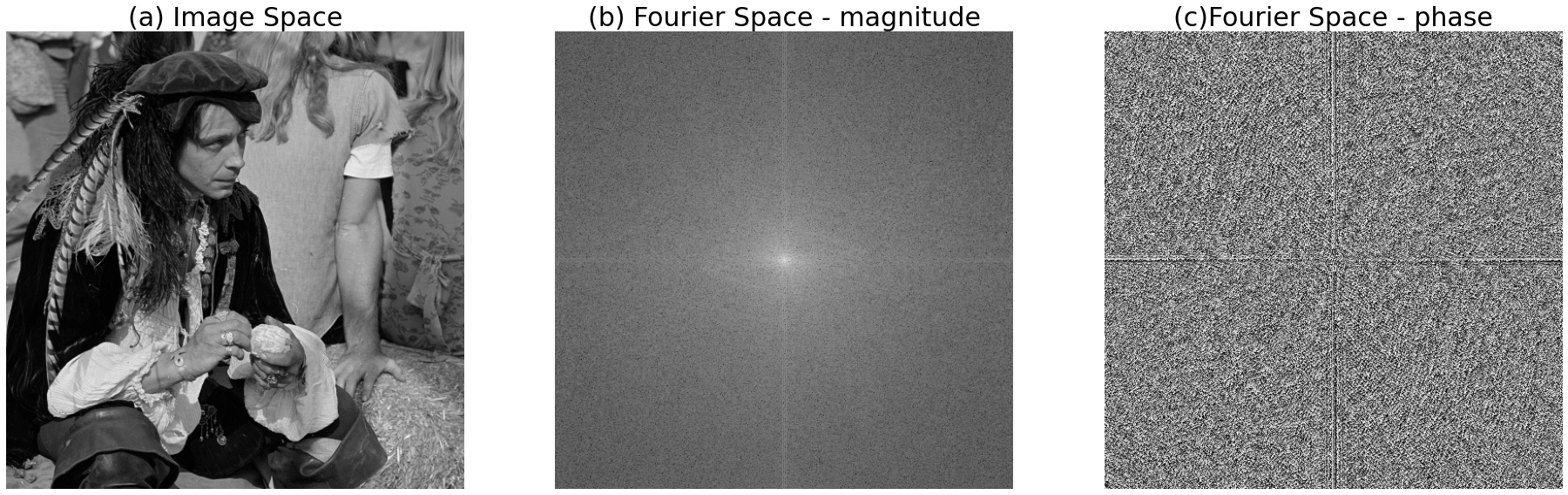}
    \caption{(a)Signal in spatial domain, (b)logarithm of the magnitude of the signal in Fourier domain, (c)phase in Fourier domain.}
    \label{fig:Fourier_Transform}
\end{figure}

 In image space we are typically interested in images that can be represented by real intensity values whereas in Fourier space the image is represented by an $M \times N$ complex-valued signal. In the BIFS framework, the components of the signal in Fourier space are modeled via the magnitude and phase; for the sake of simplicity, whenever we refer to a ``signal'' then this will indicate the Fourier space map of vectors consisting of the magnitude and phase at each (Fourier space) location. It is preferable to work with polar coordinates in the BIFS framework because, in contrast to the real and imaginary components, it is more straightforward to model the image characteristics that way. We note that the signal is typically represented such that zero frequency is represented at the center of Fourier space such that higher frequencies are towards the edges. In Figure~\ref{fig:Fourier_Transform} we illustrate an image and its corresponding Fourier signal in terms of the magnitude and the phase components. 
\par For smoothing purposes in image space, lower frequencies contain more valuable information. However, if someone is interested in enhancing edges or the high frequency structure then relevant information can be found on medium or high frequencies. The use of magnitude and phase equips us with a more convenient and intuitive way to define our prior beliefs as will be shown later. 
\par We denote as $\mathcal{F}x$ the true Fourier domain signal and $\mathcal{F}y$ the noisy Fourier domain signal. Then the posterior distribution of the signal in Fourier space is given by the formula,
\begin{equation}
    \pi(\mathcal{F}x | \mathcal{F}y) \propto \pi(\mathcal{F}x | \mathcal{F}y) \pi(\mathcal{F}x).
    \label{eq2}
\end{equation}
\par The key property that this method uses is to adopt an independence assumption for describing signals across Fourier space locations. That this assumption is reasonable stems from the spectral representation theorem \cite{subba2018statistical}\cite{bandyopadhyay2009asymptotic}; assuming that the noisy image $x$ can be characterized as a stationary process then its Fourier Transform $\mathcal{F}x$ is a decomposition of $x$ into a sum of sines and cosines with uncorrelated random coefficients (i.e., $f(m,n)$ from Equation~\ref{dft}). Because with the BIFS method we consider independence for the Fourier coefficients, we are able to break the original multidimensional image analysis problem, into a set of two-dimensional ones (each Fourier space location joint estimation of the magnitude and the phase is estimated independently of all other Fourier space locations). For example, by independently applying Bayes Theorem to each location of Fourier space and obtaining maximum a posteriori estimates of magnitude and phase, we can easily calculate the MAP estimate. Finally, we can reconstruct in image space by applying the Inverse Fast Fourier Transform (IFFT) to the MAP estimate in Fourier space.
\par The prior distributions can now be defined with respect to the independence assumptions across Fourier space locations. These distributions are characterised by a set of parameters, which we model with parameter functions. We assume a parameter $\lambda_{(k,l)}$ at the location $(k,l)$ of Fourier space and then we define a parameter function $g_\lambda$ such that, 
\begin{equation}
    g_\lambda(k,l) = \lambda_{(k,l)}.
    \label{parameter function}
\end{equation}
\par In general, the parameter function represents a multivariate function that is applied over the transformed image in Fourier space for each parameter of the prior distributions. We can let the model to be more flexible by allowing different distributions with different associated parameters for the magnitude and phase at each location $(k,l)$. Each Fourier space location is modeled by two variables, i.e.,
the magnitude and phase, which can be treated a priori as independent. So, the parameter function would essentially be a set of two dimensional functions. The magnitude is more related to the intensity of the image and the phase is related to the location of the objects. We can treat these two components independently unless we have reasonable evidence indicating the opposite. The choice of prior distribution for the magnitude can be determined from any continuous distribution that takes non-negative values. In case of a negative value, this will imply a change in the angle in the complex plane. The vector will be rotated by $\pi$ on the complex circle, so the phase will not provide the desired representation of our beliefs. Lastly, the prior for the phase can be selected by any continue distribution in the range $(-\pi, \pi]$.

To date, the BIFS
method has focused on calculating the MAP estimate. As mentioned above, this is achieved by separately finding the posterior mode at all Fourier space locations and then taking the IFFT of the combined set of Fourier space posterior modes; taking advantage of the spectral representation theorem we discussed before. The novelty of this paper is to extend the approach to enable full sampling from the posterior distribution via a new Markov Chain Monte Carlo algorithm in Fourier space. The proposed algorithm is based on the Metropolis-Hastings algorithm \cite{gelman2013bayesian}. Instead of providing just a single estimate in each Fourier space location (e.g. MAP estimate), we are therefore aiming to explore the full posterior density distribution. Thus, we will be able to obtain parameter functionals, to estimate, for instance, different quantiles and investigate how these relate to uncertainty in the reconstruction in image space through construction of Bayesian credible intervals.
\par The remainder of this paper is organized as follows. In Section~\ref{MCMC approach for BIFS} we providing the theoretical framework of our MCMC algorithm on BIFS. Next, in Section~\ref{Posterio simulation for BIFS} we apply the proposed algorithm to semi-simulated MRI data. In addition, we take advantage of the posterior sample we generate with the MCMC algorithm by constructing a posterior map in image space to detect changes in tumor across time. Lastly, in Section~\ref{Discussion and conclusions} we summarize the conclusions of this paper and we present our ideas on feature work.

\section{MCMC approach for BIFS}\label{MCMC approach for BIFS}
The quantities of interest for estimation and inference at each Fourier space location are the posterior magnitude $\rho \in \mathbb{R}^{+} \cup \{0\}$ and posterior phase $\theta \in (-\pi, \pi]$ of the signal. In addition, because we adopt the assumption of independence across Fourier space locations, we may focus on developing MCMC for just a single point on that space. In that scenario, the problem reduces to just two dimensions (real and imaginary (Cartesian); or magnitude and phase (polar)). With the independence specification across Fourier space locations
, the posterior simulation process can be applied in parallel across Fourier space locations. 
\par Our MCMC algorithm for each Fourier space location make use of both the real/ imaginary and magnitude/phase coordinate specifications of the signal. We denote $a$ and $b$ as the real and imaginary part of the observed signal, respectively, and $\alpha$ and $\beta$ as the corresponding parts of the true signal. Similarly, we define $r$ and $\psi$ to be the observed values of magnitude and phase, respectively, and $\rho$ and $\theta$ the true unknown values of the corresponding polar pair. Our goal is to generate estimates for $\alpha$ and $\beta$ where the a priori important characteristics are enhanced. The a priori important characteristics are defined through the specification of the prior for the magnitude and phase.
Thus, we can write that $a = \alpha + \epsilon_1 = \rho \cos(\theta) + \epsilon_1$ and $b = \beta + \epsilon_2 = \rho \sin(\theta) + \epsilon_2$, where $\epsilon_i, \; i = 1,2$, is additive noise with density function $\mathcal{D}$. Then making a change of variables from Cartesian to polar coordinates we have, $r = \sqrt{a^2 + b^2}$ and $\psi = \tan^{-1}(b/a)$. So, the joint distribution of $r$ and $\psi$ is given by the transformation formula, i.e.,
\begin{equation}
    \pi(r, \psi | \rho, \theta, \sigma ) = \pi (a,b|\rho, \theta, \sigma) \cdot |J|_{(a,b) \rightarrow (r, \psi)}
    \label{transformation.formula}
\end{equation}
\noindent where $|J|_{(a,b) \rightarrow (r, \psi)} = r$ is the determinant of the Jacobian matrix for the transformation.
\par From Bayes Theorem the joint posterior distribution of the magnitude and phase takes the following form,
\begin{equation}
    \pi(\rho,\theta | r, \psi, \sigma) \propto \pi(r, \psi | \rho, \theta, \sigma) \pi(\rho, \theta)
    \label{eq4}
\end{equation}
\noindent where $r$ is the observed magnitude, $\psi$ is the observed phase and $\sigma$ is the scale parameter of the likelihood distribution (detailed description is given below). We model the magnitude and phase to be a priori independent, because we do not have information about their relation. Based on that, the joint prior becomes $\pi(\rho, \theta) = \pi(\rho)  \pi(\theta)$.
\par The Metropolis-Hastings algorithm will be used to generate samples from the posterior. However, we do not propose steps based on proposal distributions for the parameters of interest directly. The reason for not doing so is that it can be seen that the marginal posterior for the phase depends on the value of magnitude, i.e., $\pi(\theta|\rho,\psi, \sigma)$. Because of the dependency of these two components, it becomes challenging to find straightforward proposal distributions, in particular the marginal distribution for the phase is difficult to work with. Our alternative approach is to propose new values with respect to the Cartesian pair instead of the polar. We already discussed that the real and imaginary values are independent and identically distributed.
In this way, it is easier to propose values for this pair of coordinates in the Cartesian framework.
\par We denote  $\mathbf{C}^{*} = (\alpha^{*}, \beta^{*})$ as the proposed values of real and imaginary part at each step. Next, we transform these values to polar coordinates, so that we have $\mathbf{P}^{*} = (\rho^{*}, \theta^{*})$. Now, we define an arbitrary (but easy to simulate from) transition probability $q(\mathbf{C},\mathbf{C}^{*})$ (i.e.,
transition probability density of moving from $\mathbf{C}$ to $\mathbf{C}^{*}$). We require that the support of $q$ is the same as the target's distribution, so we can explore all areas for which the target has positive probability (irreducibility property). Recall that the support of each Cartesian parameter is $\mathbb{R}$. But, since the acceptance probability ratio is specified in terms of polar coordinates, the corresponding transition probability is also given via the transformation formula, i.e., $q(\mathbf{P},\mathbf{P}^{*}) = q(\mathbf{C},\mathbf{C}^{*})\cdot |J|_{\mathbf{C} \rightarrow \mathbf{P}}$. The transformation formula guarantees that the support of $q$ with respect to the polar coordinate representation of the chain moves is $\mathbb{R}^{+} \cup \{0\} \times (-\pi, \pi]$ (i.e., $\mathbb{R}^{+} \cup \{0\}$ for the magnitude and $(-\pi, \pi]$ for the phase). 
\par In this approach our Metropolis-Hastings algorithm produces four chains, one for each of the magnitude, phase, real and imaginary parts. For each iteration we interchange the values from Cartesian coordinates to polar ones and vice versa and then update all the chains by the end of each iteration. We highlight that the acceptance probability is being calculated with the polar coordinates, so the Jacobian must embed on it. For the proposal step, we can take advantage of the independence between the real and imaginary parts of the signal and independently generate one value on the real-axis and one on the imaginary-axis of the complex plane from the same distribution. The proposed complex point is then transferred to polar coordinates. In this way, the difficulty of proposing a value for the phase is avoided. Algorithm~\ref{alg:JointMCMC} sets out the process for posterior sampling at each Fourier space point in detail.
\begin{algorithm}[ht]
	\SetAlgoLined
	$\mathrm{\mathbf{Initialize}}$ $\mathbf{P}^{(0)} = (\rho^{(0)}, \theta^{(0)})$;\\
	$\mathrm{\mathbf{Transform}}$ to Cartesian coordinates: $\mathbf{P}^{(0)} \rightarrow \mathbf{C}^{(0)} = (\rho^{(0)} \cos\theta^{(0)}, \rho^{(0)} \sin\theta^{(0)})$;\\
	
	\For{$t = 1, \dots, T $}{
	    $\mathrm{\mathbf{Generate}}$ proposed value  $\mathbf{C}^{*} = (a^{*}, b^{*})$ from $q(\mathbf{C}^{(t-1)}, \cdot)$\\
		$\mathrm{\mathbf{Transform}}$ to polar coordinates: $\mathbf{C}^{*} \rightarrow \mathbf{P}^{*} = \bigl ( \sqrt{a^{*2} + b^{*2}}, \tan^{-1}(b^{*}/ a^{*}) \bigr )$\\

		\texttt{\\}
        Let $\mathbf{G} = (r,\psi,\sigma)$.
	   $\mathrm{\mathbf{Calculate}}$ the joint probability of acceptance:\\
		\texttt{\\}
  
	    $A = A(\mathbf{P}^{*}, \mathbf{P}^{(t-1)}) = \min \biggl (1, \frac{\pi(\mathbf{P}^{*}|\mathbf{G})}{\pi(\mathbf{P}^{(t-1)}|\mathbf{G})} \frac{\pi(\mathbf{P}^{*})}{\pi(\mathbf{P}^{(t-1)})}
     \frac{q(\mathbf{C}^{(t-1)}, \mathbf{C}^{*})}{q(\mathbf{C}^{*}, \mathbf{C}^{(t-1)})} \frac{|J|_{\mathbf{C^{(t-1)}} \rightarrow \mathbf{P}^{(t-1)}}}{ |J|_{\mathbf{C}^{*} \rightarrow \mathbf{P}^{*}}} \biggr )$;\\
		\texttt{\\}

		$\mathrm{\mathbf{Generate}}$ $u \sim \mathrm{Unif}(0,1)$;\\
		\texttt{\\}

		\uIf{$u < A$}{$\mathbf{P}^{(t)} = \mathbf{P}^{*}$;}
		\uElse{$\mathbf{P}^{(t)} = \mathbf{P}^{(t-1)}$;}
		\texttt{\\}

		$\mathrm{\mathbf{Transform}}$ to Cartesian coordinates: $\mathbf{P}^{(t)} \rightarrow \mathbf{C}^{(t)} = (\rho^{(t)} \cos\theta^{(t)}, \rho^{(t)} \sin\theta^{(t)});$\\
		\texttt{\\}

        $\mathrm{\mathbf{Update}}$ chains $\mathbf{P, C}$
	}
    \caption{Joint Markov Chain Monte Carlo}
    \label{alg:JointMCMC}
\end{algorithm}

\section{Posterior simulation for BIFS}\label{Posterio simulation for BIFS}
In this section we will describe and illustrate the specific MCMC approach under consideration, i.e., the definition of the likelihood and the prior distributions under the assumption that the noise in image space is independently and identically distributed (i.i.d.) Gaussian (i.e., complex Gaussian noise in Fourier space). So, following the notation we described in the previous section, the observed real and imaginary parts of the signal can be defined as, $a = \alpha + \epsilon_1$ and $b = \beta + \epsilon_2$, where $\epsilon_i \sim N(0, \sigma^2_i)$ for $i=1,2$, respectively. Moreover, we will examine mixing and convergence of Markov chains for the magnitude and phase at different locations in Fourier space as well as generate maps of acceptance probabilities over Fourier space. 
\par Firstly, specify the modelling framework beginning with the likelihood distribution. After applying some calculations with trigonometric identities to Equation~\ref{transformation.formula}, we obtain the joint likelihood distribution \cite{rowe2004complex} for the observed magnitude and phase,
\begin{equation}
    \pi(r, \psi | \rho, \theta, \sigma ) = \frac{r}{2\pi \sigma^2} \exp \biggl ( - \frac{1}{2\sigma^2} \bigl [ r^2 + \rho^2 - 2\rho r\cos(\psi - \theta) \bigr ] \biggr) 
\end{equation}
\noindent where $r$ is the observed value of the magnitude, $\rho$ is the true value of the magnitude, $\psi$ is the observed value of the phase, $\theta$ is the true value of the phase
and $\sigma$ is the standard deviation of the real and imaginary Gaussian noise components of the independent and identically distributed
signal in Fourier space. We define $\sigma_{\mathrm{noise}}$ to be the standard deviation of the Gaussian noise in image space. Therefore, the parameter $\sigma$ can be estimated as, $\hat{\sigma} = \frac{\sigma_{\mathrm{noise}}}{\sqrt{2}}$.
The quantity $\sigma_{\mathrm{noise}}$ can be estimated by a flat patch of noise in image space, i.e., a region of the image that consists only of noise and contains no information about the contents of the image.
\par The prior distribution for each Fourier space location will be defined via parameter functions of the form given in Equation~\ref{parameter function}. 
As we discussed in Section~\ref{Introduction}, we will treat the priors for magnitude and phase independently.
For the prior specification of the magnitude, we use a truncated Normal prior
, $\rho \sim TN(\mu_{\nu}, \sigma^{2}_{\nu})$, where $\nu=(k,l)$, with support on $[0, \infty)$. The reason for choosing truncated normal as a prior for the magnitude is that it allow as to model the mean $\mu_\nu$ and the variance $\sigma^2_\nu$ with separate independent parameter functions. However, in this section we set the mean parameter $\mu_\nu = \mu(k,l)$ to be proportional to the standard deviation of the truncated normal in the magnitude prior, i.e., $\sigma_\nu=\sigma(k,l)$
, i.e., $\mu_\nu = c \sigma_\nu$, where $c$ is a user specified constant.  The specification of parameter values $\mu_\nu, \sigma_\nu$ over Fourier space locations is based on the parameter function of the form given in Equation~\ref{parameter function}. A practical way to define the parameter function of the magnitude for each location $(k,l)$, is based on its distance from the origin of Fourier space. A reasonable suggestion for the parameter function is to be reciprocal to the distance from $(0,0)$. More specifically, we define
\begin{equation}
    g_\lambda(k,l) = \frac{\lambda}{(k^2 + l^2)^{d/2}} = \frac{\lambda}{|\nu|^d}. 
    \label{definitionparameterfunction}
\end{equation}
\noindent where $\lambda>0$ and $d\in \mathbb{R}$ are predetermined parameters and $|\nu|=\sqrt{k^2 + l^2}$ is the Euclidean distance of $(k,l)$ from the origin.
Similarly, in a three dimensional setting the denominator of Equation~\ref{definitionparameterfunction} will be $|\nu| = \sqrt{k^2 + l^2 + z^2}$, where $z$ is the coordinate of the third dimension of the 3D-Euclidean space.
\par The prior distribution for the phase is more difficult to be determined. For example, the displacement of an object in the image, significantly affects the phase. Since it is hard to know in advance the exact position of the objects within an image, a non-informative prior can be considered. Therefore, we use a uniform prior on the unit circle for the phase, $\theta \sim U(-\pi, \pi)$.
\par Given the set up above for the likelihood and prior at each point in Fourier space, we can now specify the joint Metropolis-Hastings algorithm for $\rho$ and $\theta$ based on Algorithm~\ref{alg:JointMCMC}. The first step is to propose values $\mathbf{C}^{*}$, for the real and the imaginary parts. The proposal distribution we use for each of these parameters is the Normal with mean the current value and variance $\xi_k$ that can be different for each Fourier space location. Because of this normal (symmetric) proposal step the acceptance probability ratio of the acceptance probability simplifies to a Metropolis sampler,
\begin{align*}
    A(\mathbf{P}^{*}, \mathbf{P}^{(t-1)}) &= \min \biggl (1, \frac{\pi(\mathbf{P}^{*}|\mathbf{G})}{\pi(\mathbf{P}^{(t-1)}|\mathbf{G})} \frac{\pi(\mathbf{P}^{*})}{\pi(\mathbf{P}^{(t-1)})}
     \frac{q(\mathbf{C}^{(t-1)}, \mathbf{C}^{*})}{q(\mathbf{C}^{*}, \mathbf{C}^{(t-1)})} \frac{|J|_{\mathbf{C^{(t-1)}} \rightarrow \mathbf{P}^{(t-1)}}}{ |J|_{\mathbf{C}^{*} \rightarrow \mathbf{P}^{*}}} \biggr )\\
    &= \min \biggl (1,  \frac{\pi(\mathbf{P}^{*}|\mathbf{G})}{\pi(\mathbf{P}^{(t-1)}|\mathbf{G})} \frac{\pi(\mathbf{P}^{*})}{\pi(\mathbf{P}^{(t-1)})} \frac{\rho^{(t-1)}}{\rho^{*}} \biggr )\\
    &= \min \biggl (1, \exp \biggl (  \bigl ( \rho^{(t-1)} - \rho^{*} \bigr ) \Gamma + \Beta(\theta^{*}) \rho^{*} - \Beta(\theta^{(t-1)})\rho^{(t-1)} \biggr ) \biggr )
    \numberthis
    \label{eq8}
\end{align*}
\noindent where $\Gamma = \frac{\sigma_k^2 + \sigma^2}{2\sigma_k^2 \sigma^2}$, $\Beta(\theta) = \frac{r}{\sigma^2}\cos(\psi - \theta) + \frac{\mu_k}{\sigma_k^2}$ and the Jacobian of the transformation from Cartesian to polar coordinates is $|J|_{\mathbf{C} \rightarrow \mathbf{P}} = \rho$, where $\rho$ is the value of magnitudes' chain contained in $\mathbf{P}$. The acceptance probability of the candidate pair $\mathbf{P^{*}}$ is easy to compute by simply taking its logarithm. Subsequently, the logarithm of the generated uniform value $u$ will be taken into account instead. 

\subsection{Simulated brain application}
\begin{figure}
    \centering
    \includegraphics[width=15cm, height=7cm]{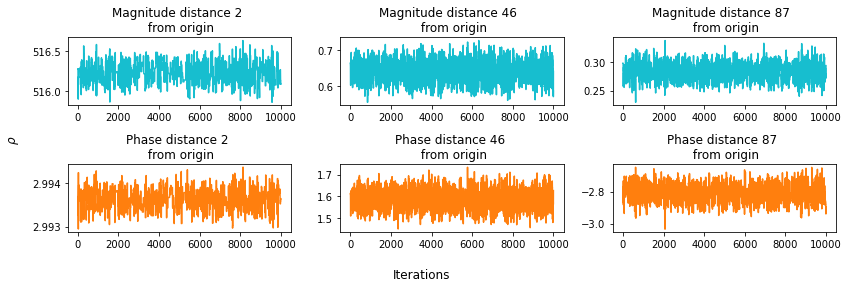}
    \caption{Traceplots for magnitude and phase that correspond to close, middle distance and far from the origin of Fourier space respectively.}
    \label{fig:traceplots}
\end{figure}

\begin{figure}
    \centering
    \includegraphics[width=11cm, height=10cm]{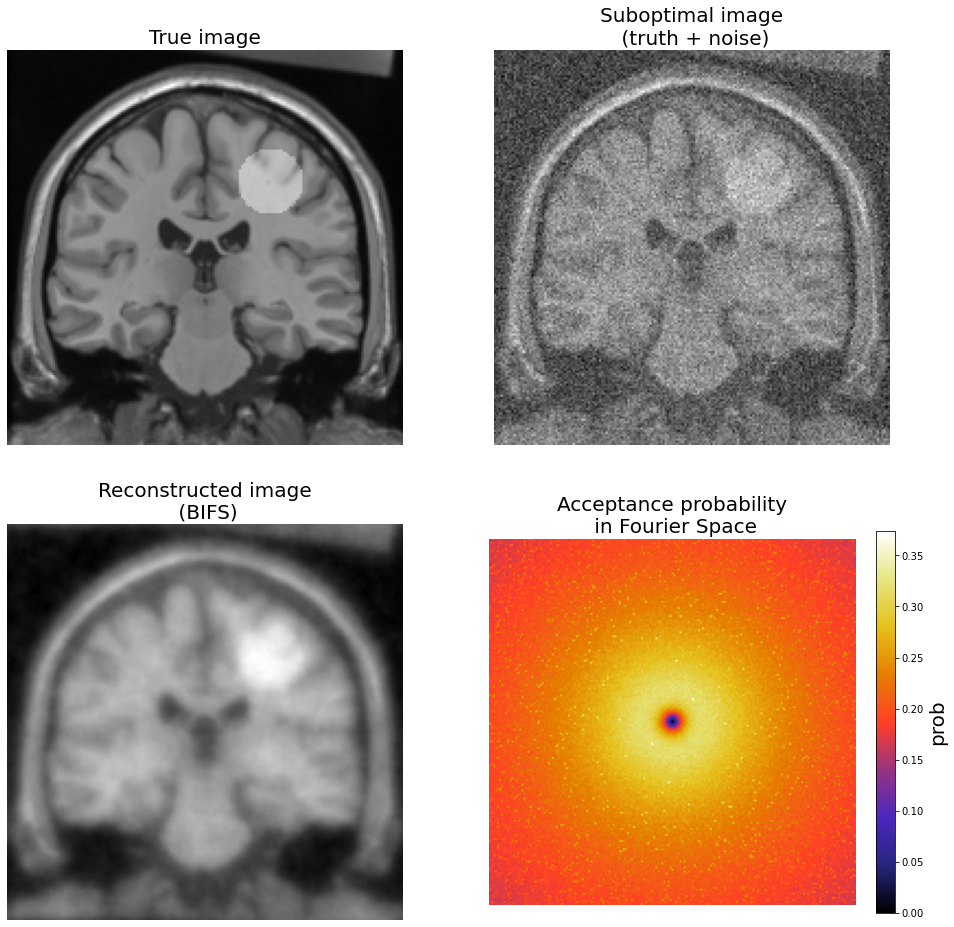}
    \caption{Reconstruction results with MCMC BIFS.}
    \label{fig:reconstruction}
\end{figure}

The data we use to test the performance of our algorithm comes from a single slice of an average of 27 T1-weighted MRI scans of the same individual \cite{collins1998design}. The data are available from Montreal Neurological Institute (MNI), McGill University, McConnell Brain Imaging Center web site\footnote{https://nist.mni.mcgill.ca/atlases/}. The dimension of the selected slice is $181\times181$. We manually place a circular tumor in the brain by increasing the intensity values of a circular set of pixels. The elevated circle can clearly be seen in the top left quadrant in the Figure~\ref{fig:reconstruction}. Note that the true image is given solely for the purpose of comparison with the result of the algorithm. In practise, it will not be available. To form a degraded dataset we add i.i.d Gaussian noise to the image. We are  trying to mimic conditions that occur in other fast imaging modalities (e.g. DWI and fMRI) but by no means claim that this would be a typical realization of either modality. In Figure~\ref{fig:traceplots} we illustrate the example chains of the magnitude and phase for three different locations in Fourier space.
\par The reconstruction of the noisy image as well as the acceptance probability map that we obtain after applying our method in Fourier space are presented in Figure~\ref{fig:reconstruction}. Note that, the parameter function we use in this example is that of Equation~\ref{definitionparameterfunction}, where $\lambda=1$ and $d=1$. We notice that the quality of the noisy image is significantly improved in the reconstructed image (Figure~\ref{fig:reconstruction}) and some features have become more distinct. The artificial tumor is visually quite degraded in the presence of the additive noise, but our algorithm has managed to enhance it. Finally, one additional note is that it can be seen in the bottom right panel of Figure~\ref{fig:reconstruction} that based on having a constant variance for the proposal step the acceptance probability varies over Fourier space. Ideally, we are aiming for an acceptance probability around 0.234 across all Fourier space locations which has been proposed as an asymptotically optimal acceptance rate \cite{gelman1997weak} for the exploration of the target distribution. The actual range of the aforementioned quantity we achieve mostly varies from around 0.17 to 0.35, which is a reasonable range for the acceptance probabilities. However, in the center of Fourier space we do observe some smaller values. Since we are applying MCMC independently across Fourier space locations it is straightforward to either manually adjust the variance of the proposal distribution in those locations to increase the acceptance rate, or even to define a parameter function for proposal step size over Fourier space that will lead to an approximately flat acceptance rate map around the desired rate. Note that of course any samples acquired with any adaptation phase should be discarded.
\par
\begin{figure}
    \centering
    \includegraphics[width=15cm, height=10cm]{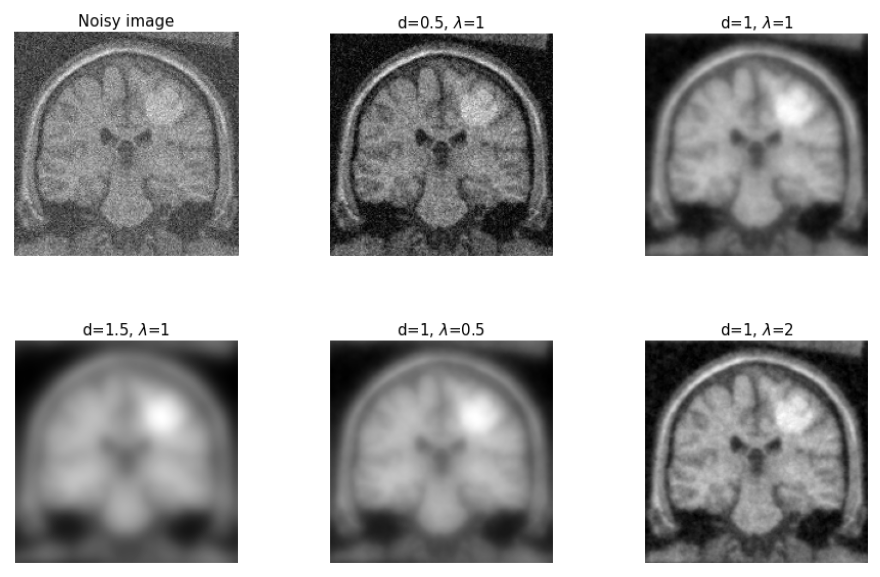}
    \caption{Results for different parameter functions}
    \label{fig:parameter_funs_plots}
\end{figure}

 In practice, we find that the parameter function is the most influential factor of model specification for the image reconstruction/enhancement characteristics of posterior estimation. For example, Figure~\ref{fig:parameter_funs_plots} shows the effect of just changing the parameters $\lambda$ and $d$ of the parameter function in Equation~\ref{definitionparameterfunction}. As $d$ increases the prior distribution decays faster to zero as we move away from the origin of Fourier space. That gives a priori less probability to higher frequencies and therefore smoother posterior estimates. The parameter $\lambda$ affects the function in a reciprocal way.
\subsection{MCMC samples for posterior change mapping}
MCMC sampling allows the generation of a wide range of posterior distribution summaries. For example MCMC can be used to generate a probability map to detect changes in tumor state during successively acquired images. In this example, we simulate two images with two different realizations of i.i.d Gaussian noise (the same SD for each image), with the only other difference between the two images being the intensity or/and location of the tumor. Without loss of generality we calculate the probability map $\textbf{\Lambda}$ that the intensity of the tumor in the second image has increased comparing to the first one. The MCMC algorithm is applied separately to each image (i.e. treating them as independent images). Then, we obtain the posterior distribution samples for each timepoint image for all states $t$, where $t=1,2,\dots,T$, of the Markov chains. We first combine the Markov chains across Cartesian coordinates into one 3D array. Applying inverse Fourier transform to the 2D sub-array at each state in the Markov chain reflects a posterior image sample from the Markov chain.
\par We define $\textbf{s}^{(t)}_{m,n}$ to be the first image and $\textbf{v}^{(t)}_{m,n}$ the second image at state $t$ of Markov chains, where $m,n=1,2,\dots,181$. We take the difference at each state of the chain $t$, giving $\textbf{\Delta}^{(t)}_{m,n} = \textbf{v}^{(t)}_{m,n} - \textbf{s}^{(t)}_{m,n}$. For each $\textbf{\Delta}_{m,n}$ we compute the probability map $\textbf{\Lambda}(m,n)$ where its value, at location $(m,n)$, is the proportion of samples above zero across the $T$ states, i.e.,
\begin{equation}
    \textbf{\Lambda}(m,n) = 1/T\sum_{t=1}^{T} I(\textbf{\Delta}^{(t)}_{m,n} > 0).
\end{equation}
\begin{figure}[h]
    \centering
    \includegraphics[width=15cm, height=10cm]{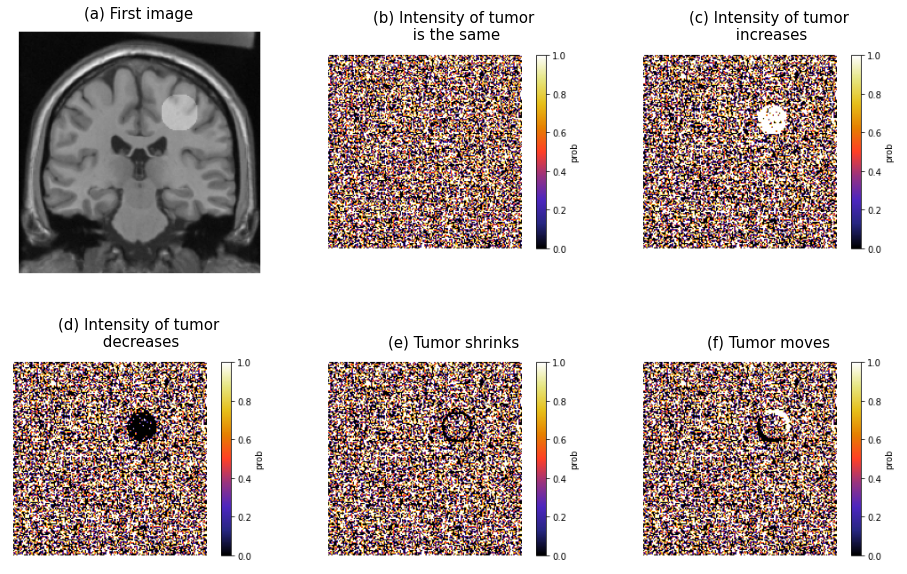}
    \caption{(a) The suboptimal first image $\textbf{s}$. The probability map $\textbf{\Lambda}$ when (b) the two sequential images are identical, (c) the intensity of tumor slightly increases in second image, (d) the intensity of tumor slightly decreases in second image, (e) the tumor shrinks in second image, (f) the tumor moves up and right in the second image.}
    \label{fig:summary}
\end{figure}

\par More specifically, for a pixel $(k,l)$ that is outside the tumor region in both the first and second image, we expect the value of the probability $\textbf{\Lambda}(k,l)$ to be on average equal to 0.5. Therefore, if the intensity of the tumor in the second image has been increased then we anticipate that the probability map $\textbf{\Lambda}$ will have values close to one in the region of the tumor. On the other hand, if the intensity of the tumor has been decreased then we expect the probability map at that region to have values close to zero. In addition, if the intensity of the tumor is the same for the two images then every pixel will have on average 0.5 probability. In the case where the tumor has been slightly moved between the first and second image then we anticipate that the locations where the tumor is present in the first image but not present in the second to have on average values of $\textbf{\Lambda}$ close to zero. In contrast, regions where the tumor is present in second image but not in the first, we expect them to have on average values of $\textbf{\Lambda}$ close to 1. In Figure~\ref{fig:summary}, we illustrate the probability map $\textbf{\Lambda}$ examining various alternatives by changing the intensity and/or location of the tumor.

\section{Discussion and conclusions}\label{Discussion and conclusions}

In this paper we generalize the BIFS method by extending the previous MAP estimation approach to fully sampling from the posterior distribution with MCMC. By exploiting the assumption of independence across Fourier space locations, we manage to break a multidimensional MCMC problem in image space into a large set of two dimensional problems in Fourier space (magnitude and phase are the parameters of interest on each signal). As a result, one of the advantages of the proposed method is that it is computationally relatively inexpensive.
\par To date BIFS method has relied on generating just the Maximum a Posterior estimate for each location in Fourier space. The novelty of the proposed approach is the ability to sample from the posterior distribution and the ability to generate of a range of estimates and quantiles from the posterior distribution. In this paper, we highlight the flexibility of using MCMC samples by generating posterior probability maps that can monitor the progression of a tumor in the brain. In the future, we plan a more optimal approach to this problem that will utilize the longitudinal nature of the image acquisitions rather than treating each image timepoint independently. 
\par One limitation of BIFS in general (and therefore our MCMC formulation of BIFS) is the assumption of stationarity in image space wrapped on a torus. We assume that every frequency has zero mean (except at the origin). In practise, for many medical images applications the region of interest is located at the center of the image and not near the edges, so it will be minimally affected by bias caused by differential intensities at the edges. Furthermore, the edges of images are often outside of the body tissue and therefore the underlying intensity mean in most regions at the edges do have intensity zero. A conventional ad-hoc approach to address this general problem, when it does exist, is to extend the perimeter of the image by padding with pixels of e.g. local mean intensity. The subject of our future work will be to explore models for non-stationarity conditions under a transformation framework that does not require stationarity by using wavelets.

\section*{Acknowledgements}

Research reported in this manuscript was supported by National Institute of Biomedical Imaging and Bioengineering of the National Institutes of Health under award number R01EB022055.


\printbibliography








\end{document}